\documentclass[journal,article,submit,moreauthors,pdftex,10pt,a4paper]{mdpi} 
\firstpage{1} 
\articlenumber{x}
\doinum{10.3390/------}
\pubvolume{xx}
\pubyear{2018}
\copyrightyear{2018}
\history{Received: date; Accepted: date; Published: date}

\Title{Hard-core Radius of Nucleons  within  the Induced Surface Tension Approach}
\Author{K. A. Bugaev$^{1}$, A. I. Ivanytskyi$^{1,2}$, V. V. Sagun$^{1,3}$, 
B. E. Grinyuk$^{1}$, D. O. Savchenko$^{1}$,	G. M. Zinovjev$^{1}$,
E. G. Nikonov$^{4}$, L. V.  Bravina$^{5}$,	E. E.  Zabrodin$^{5, 6, 7}$,
D. B. Blaschke$^{7, 8, 9}$,	A. V.  Taranenko$^{7}$,	L. Turko$^{8}$
}

\AuthorNames{K. A. Bugaev, A. I. Ivanytskyi, V. V. Sagun, 
	B. E. Grinyuk, D. O. Savchenko,	G. M. Zinovjev,
	E. G. Nikonov, L. V.  Bravina,	E. E.  Zabrodin,	D. B. Blaschke,	A. V.  Taranenko,	L. Turko}

\address{%
	$^{1}$ \quad Bogolyubov Institute for Theoretical Physics of the National Academy of Sciences of Ukraine, 03680 Kiev,  Ukraine
		\\
	$^{2}$ \quad Department of Fundamental Physics, University of Salamanca, 
		Plaza de la Merced s/n 37008, Spain
		\\
	$^{3}$ \quad Centro de Astrof\'{\i}sica e Gravita\c c\~ao  - CENTRA,
		Departamento de F\'{\i}sica, Instituto Superior T\'ecnico,
		Universidade de Lisboa, 1049-001 Lisboa, Portugal
		\\
	$^{4}$ \quad Laboratory for Information Technologies, Joint Institute for Nuclear Research,   Dubna 141980, Russia
		\\
	$^{5}$ \quad Department of Physics, University of Oslo, PB 1048 Blindern, N-0316 Oslo, Norway 
		\\
	$^{6}$ \quad Skobeltzyn Institute of Nuclear Physics, Moscow State University, 119899 Moscow, Russia
		\\
	$^{7}$ \quad National Research Nuclear University ``MEPhI'' (Moscow Engineering Physics Institute), 115409 Moscow, Russia
		\\
	$^{8}$ \quad Institute of Theoretical Physics, University of Wroclaw, pl. M. Borna 9, 50-204 Wroclaw, Poland
		\\
	$^{9}$ \quad Bogoliubov Laboratory of Theoretical Physics, JINR Dubna, Joliot-Curie str. 6, 141980 Dubna, Russia
	}

\corres{Correspondence: $^*$bugaev@fias.uni-frankfurt.de}

\abstract{
In this work we discuss a novel approach to model the hadronic and nuclear matter equations of state 
using  the induced surface tension concept. 
Since the obtained equations of state, classical and quantum, are among the most
successful ones in describing the properties of low density phases of strongly interacting matter,
they set  strong restrictions on the possible value of the hard-core radius of nucleons. 
Therefore,  we perform a detailed analysis of its value which follows from hadronic and nuclear matter properties and find the most trustworthy range of its values: the hard-core radius of nucleons is  0.3--0.36 fm.
A comparison with the phenomenology of neutron stars implies that the  hard-core radius of nucleons has to be temperature and density dependent.
}
\preto{\abstractkeywords}{\nolinenumbers}
\begin{document}
\section{Introduction}
\label{intro_Bugaev_talk}

A reliable and precise  determination of  major characteristics of symmetric  nuclear matter is of fundamental importance \cite{Horst86,Lattimer12,StellarSMM13,David15,Dutra14} 
not only for the nuclear spectroscopy and for nuclear physics of intermediate energies, but also for nuclear astrophysics in view of possible phase transformations which may occur in compact astrophysical objects such as neutron stars, and hypothetical hybrid and quark stars.  
Such characteristics of infinite nuclear matter as the normal density $n_0 \simeq 0.16$ fm$^{-3}$ at zero pressure and zero temperature, its binding energy per nucleon $W_0 =-16$ MeV and its incompressibility factor $K_0\simeq 250-315$ MeV \cite{Stone:2014wza} are of great importance
 for various phenomenological approaches,  since these characteristics  are widely used for determination of  the model parameters. 
Furthermore,  such a parameter of the nuclear  matter as the hard-core radius (HCR) of nucleons $R_N$ plays an important role not only in nuclear physics \cite{Horst86,StellarSMM13}, but  also in nuclear astrophysics  \cite{Lattimer12,David15} and in the physics of heavy ion collisions (HIC) \cite{Andronic:05,Oliinychenko:12,HRGM:13,SFO:13,Sagun,Sagun2,Stachel:2013zma,BugaevUJP16,Bugaev:2015,Bugaev:2016a}.  However, in  the literature one can find any value of $R_N$  in the range $0.3-0.7$ fm.   Partly the problem is related to the fact that  almost all equations of state  (EoS) with  the hard-core repulsion employ the Van der Waals (VdW) approximation which is applicable  only at low particle number densities.  

However, recently a novel and convenient  approach to the EoS has been developed which allows one  to safely  go beyond the VdW approximation for any number of HCR (multicomponent case) \cite{Bugaev:2017a,IST1,IST2,IST3}.  
Having a single  additional parameter compared to the multicomponent VdW EoS this approach enables us to describe on the same footing the data measured in HIC,  to reproduce the nuclear matter properties up to five normal nuclear densities and to describe the mass-radius relation of  neutron stars.  
Here we consider the  constraints which follow from the proton flow and from the  S-matrix approach, 
and discuss how they allow one to determine the most trustworthy range of  $R_N$ values. 
We draw some conclusions for developing  the EoS  of neutron star matter. 

The work is organized as follows. In Sect.~2 we recall the main equations of the  hadron resonance gas model  (HRGM) 
\cite{IST1,IST2,IST3} based on the concept of induced surface tension \cite{IST14}. 
The new results on the quantum formulation of the induced surface tension 
equation of state for nuclear matter are discussed 
in Sect.~3, whereas our conclusions are summarized in Sect.~4.

\section{ Multicomponent formulation of HRGM with hard-core repulsion}

For many years the HRGM  
\cite{Andronic:05,Oliinychenko:12,HRGM:13,SFO:13,Sagun,Sagun2,Stachel:2013zma,BugaevUJP16,
Bugaev:2015,Bugaev:2016a, Bugaev:2017a,IST1,IST2,IST3} 
is successfully  used to finding out  the parameters of chemical freeze-out (CFO) from the  hadronic 
yields measured experimentally in high energy nuclear collisions. Presently  the HRGM with multicomponent hard-core 
repulsion between hadrons 
 \cite{Oliinychenko:12,HRGM:13,SFO:13,Sagun,Sagun2,Bugaev:2015,Bugaev:2016a, Bugaev:2017a,IST1,IST2,IST3} gives  the  
most successful description of all independent hadronic multiplicity 
ratios which have been measured in the heavy ion collisions high energy experiments performed from early 70' (Bevalac) till present  over BNL-AGS, GSI-SIS, CERN-SPS, BNL-RHIC to CERN-LHC at   the broad center of mass energies $\sqrt{s}_{NN}$ from  2.7 to 5020~GeV.
There exist  three major grounds  to consider  the HRGM with multicomponent hard-core 
repulsion  as the realistic  EoS of hadronic matter at high temperatures and moderate particle number densities. 
Firstly, for a long time it is well known  that for temperatures below 170 MeV and moderate  baryonic charge densities (below about twice nuclear saturation density) the  mixture of stable hadrons and their resonances
whose interaction is described by  the quantum  second virial 
coefficients behaves almost  like  a mixture of  ideal gases of stable 
particles which, however,  includes both the hadrons and their 
resonances, but  with their averaged vacuum values of masses \cite{Raju}. As it was 
demonstrated  in Ref.~\cite{Raju}, the main physical reason for this kind of  behavior is 
rooted in   an almost complete cancellation between the attractive  and 
repulsive terms  in  the quantum second virial coefficients. 
Hence, the residual  deviation from the ideal gas (a weak repulsion) 
can be modeled  by the classical second virial coefficients. 

Secondly,  by considering   the HRGM as the hadronic matter EoS 
one can be sure that  its pressure will never  exceed the one of the quark-gluon plasma.
The latter may occur,  if the hadrons are treated as the 
mixture of ideal gases \cite{IST1,Satarov10}. 
Thirdly,  an additional  reason to regard  the HRGM as hadronic matter EoS
 in the vicinity of CFO is the practical one: 
since the hard-core repulsion is a contact interaction, 
the energy per particle of such an EoS equals to the one of the ideal gas, 
even for the case of  quantum  statistics  \cite{IST3}. 
Therefore, during the evolution of the system after 
CFO to the kinetic freeze-out  one will not face a hard mathematical problem  \cite{KABkinFO1,KABkinFO2} to somehow 
``convert'' the potential energy of interacting particles into their kinetic energy 
and into the masses of  particles which appear due to resonance decays. 

Apparently, these reasons allow one to consider the HRGM as an extension of the statistical bootstrap model \cite{Hagedorn} 
augmented 
with  the hard-core repulsion, but for  a truncated hadronic mass-volume  spectrum,  and to
effectively apply  it to the description of   hadronic multiplicities measured in 
the heavy ion collision experiments. 

Although  many valuable findings were  obtained with the HRGM during  last few years, 
at the moment  the HCR are well established for the most 
abundant hadrons only, i.e.  for pions ($R_\pi \simeq 0.15\pm 0.02$ fm), for the
lightest  K$^\pm$-mesons ($R_K \simeq 0.395\pm 0.03$ fm), for nucleons 
($R_p \simeq 0.365\pm 0.03$ fm) and for the lightest (anti)$\Lambda$-hyperons 
($R_\Lambda \simeq 0.085\pm 0.015$ fm) \cite{IST1,IST2}. 
Nevertheless, there is a confidence   that in few years from now  the new data of high quality 
which  will be   measured  at RHIC BNL 
(Brookhaven) \cite{RHIC17},  NICA JINR (Dubna) \cite{NICA} and FAIR GSI (Darmstadt) 
\cite{FAIR}, will help us to find out  the HCR of other 
measured hadrons with  unprecedentedly high accuracy. 
However,  one should remember that the traditional 
multicomponent HRGM based on the VdW approximation 
is not suited for such a purpose, since for $N\sim 100$ different HCR, where  
$N$ corresponds to the various hadronic species produced  in a collision, 
one has  to find a solution of $N$  transcendental equations. 
Therefore, an  increase of the number of  HCR to  $N\sim 100$ will lead to hard 
computational problems  for  the traditional  HRGM with  multicomponent hard-core 
repulsion. To resolve this principal problem   the new HRGM based on the 
induced surface  tension (IST)  concept  \cite{IST14} was recently developed  in 
Refs.~\cite{IST1,IST2,IST3}.

The IST EoS  is a system of two coupled  equations for the pressure $p$ and the 
IST coefficient  $\Sigma$
\begin{eqnarray}
\label{EqI}
p &=& \sum_{k=1}^N  p_k =  T \sum_{k=1}^N \phi_k \exp \left[ 
\frac{\mu_k}{T} - \frac{4}{3}\pi R_k^3 \frac{p}{T} - 4\pi R_k^2 
\frac{\Sigma}{T} \right]
\,, \\
\label{EqII}
\Sigma &=& \sum_{k=1}^N  \Sigma_k =  T \sum_{k=1}^N R_k \phi_k \exp 
\left[ \frac{\mu_k}{T} - \frac{4}{3}\pi R_k^3 \frac{p}{T} - 4\pi R_k^2 
\alpha \frac{\Sigma}{T} \right] \,,\\
\label{EqIII}
\mu_k &=& \mu_B B_k + \mu_{I3} I_{3k} + \mu_S S_k \,,
\end{eqnarray}
where $\alpha = 1.245$, and $\mu_B$, $\mu_S$, $\mu_{I3}$ are the chemical potentials of
baryon number, the strangeness, and the third projection of the isospin, respectively. 
Here $B_k$, $S_k$, $I_{3k}$, $m_k$ and $R_k$ denote, respectively, the corresponding charges, mass, and HCR of the $k$-th hadronic species. The sums in 
Eqs.~(\ref{EqI}) and (\ref{EqII}) run over all hadronic species including
their  antiparticles which  are considered as independent species.  Therefore,
 $p_k$ and $\Sigma_k$ are, respectively, the partial pressure and 
the partial induced surface tension coefficient of the $k$-th hadronic 
species. 

In  Eqs.~(\ref{EqI}) and (\ref{EqII}) the  thermal density $\phi_k$  of the $k$-th hadronic sort
contains  the Breit-Wigner mass attenuation. Hence, 
 in the Boltzmann approximation (the quantum  gases are discussed in Ref.  \cite{IST3}) it can be cast
\begin{eqnarray}
\label{EqIV}
\phi_k = g_k  \gamma_S^{|s_k|} \int\limits_{M_k^{Th}}^\infty  \,  
\frac{ d m}{N_k (M_k^{Th})} 
\frac{\Gamma_k}{(m-m_{k})^{2}+\Gamma^{2}_{k}/4} 
\int \frac{d^3 p}{ (2 \pi)^3 }   \exp \left[{\textstyle  -
\frac{ \sqrt{p^2 + m^2} }{T} }\right] \,.
\end{eqnarray}
Here $g_k$ is the degeneracy factor of the $k$-th hadronic species,
$\gamma_S$ is the strangeness suppression factor \cite{Rafelski}, 
$|s_k|$ is the number of valence strange quarks and antiquarks in this 
hadron species, and the quantity $\displaystyle {N_k (M_k^{Th})} \equiv \int
\limits_{M_k^{Th}}^\infty \frac{d m \, \Gamma_k}{(m-m_{k})^{2}+
\Gamma^{2}_{k}/4} $ denotes 
a normalization factor with $M_k^{Th}$ being the decay 
threshold mass of the $k$-th hadronic sort, while  $\Gamma_k$ denotes its   width.

To employ   the system of Eqs.~(\ref{EqI}), (\ref{EqII}), and (\ref{EqIII}) 
to an investigation   of heavy  ion  collisions one has to supplement it  by 
the strange charge conservation condition\begin{eqnarray}
\label{EqV}
n_S\equiv\frac{\partial p}{\partial \mu_S}= \sum_k S_k\, n_k=0 \,,
\end{eqnarray}
which provides  a vanishing net strange charge. 
Here $n_k$ is the particle number density of hadrons of sort $k$ defined  by the following  system
of equations
\begin{eqnarray}\label{EqVIA}
&&\hspace*{-11mm}n_k \,\,~\equiv~\, \, \frac{\partial  p}{\partial \mu_k} = \frac{1}{T} \cdot \frac{p_k \, a_{22} 
- \Sigma_k \, a_{12}}{a_{11}\, a_{22} - a_{12}\, a_{21} } \,, \quad a_{11} ~=~ 1 + \frac{4}{3}\, \pi \sum_k  R_k^3 \frac{p_k}{T} \,, \, \quad  a_{12} ~=~ 4 \pi \sum_k R_k^2 \frac{p_k}{T} \, , \quad \\
&&\hspace*{-11mm}a_{22} ~=~ 1 + 4 \pi \alpha \sum_k  R_k^2  \frac{\Sigma_k}{T} \,, \quad a_{21} ~=~ \frac{4}{3} \pi \sum_k R_k^3\frac{\Sigma_k}{T} \,.
\end{eqnarray}

In contrast to the traditional  multicomponent HRGM formulations to 
determine the particle number densities $\{ n_k \}$ one needs to 
solve only a system of  three equations, i.e.  Eqs. (1), (2) and (5), 
irrespective to the number of different HCR in the EoS. 
Hence, we believe that the IST EoS given by the system (1)-(5) is 
well  suited for the analysis of all hadronic multiplicities 
which will be measured soon  at RHIC, NICA and FAIR. 

Compared to the VdW EoS,   the IST EoS has 
another great advantage of the IST EoS, since  it is  valid  up to the 
packing fractions $\eta \equiv \sum_k \frac{4}{3}\pi R_k^3 n_k 
\simeq  0.2$ \cite{IST1,IST2,IST3}  at  which  the VdW 
approximation  employed  in the traditional HRGM \cite{Andronic:05,Oliinychenko:12,HRGM:13,SFO:13,Sagun,Sagun2} 
becomes completely  incorrect (see a discussion below).

From  the particle number density (\ref{EqVIA}) of the $k$-th species of 
hadrons one can find out  their  thermal  $N_k^{th} =V n_k$ ($V$ is the 
effective volume of CFO hyper-surface) and  total multiplicity  $N_k^{tot}$.
The total multiplicity  $N_k^{tot}$  accounts  for the hadronic decays after the CFO and,
hence, the ratio of total hadronic multiplicities at CFO can be written 
\begin{equation}
\label{EqIX}
\frac{N^{tot}_k}{N^{tot}_j}=
\frac{n_k+\sum_{l\neq k}n_l\, Br_{l\rightarrow k}}{n_j+
\sum_{l\neq j}n_l \, Br_{l\rightarrow j}}\,.
\end{equation}
Here $Br_{l\rightarrow k}$ denotes  the branching ratio, i.e., a probability 
of particle $l$ to decay strongly into a particle $k$. Further details on 
the actual fitting procedure of experimental hadronic multiplicities  by  the HRGM can be found in  
\cite{Sagun, IST1}.

\begin{figure}[t]
\centering
\mbox{\includegraphics[width=70mm,clip]{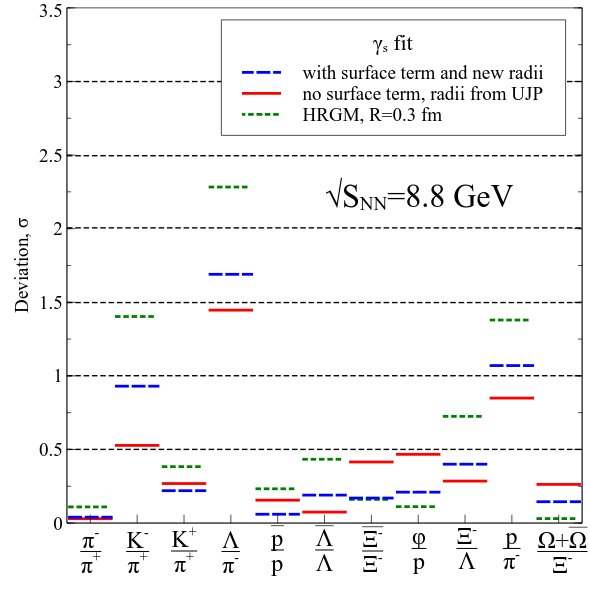}
\hspace*{0.2mm}
\includegraphics[width=70mm,clip]{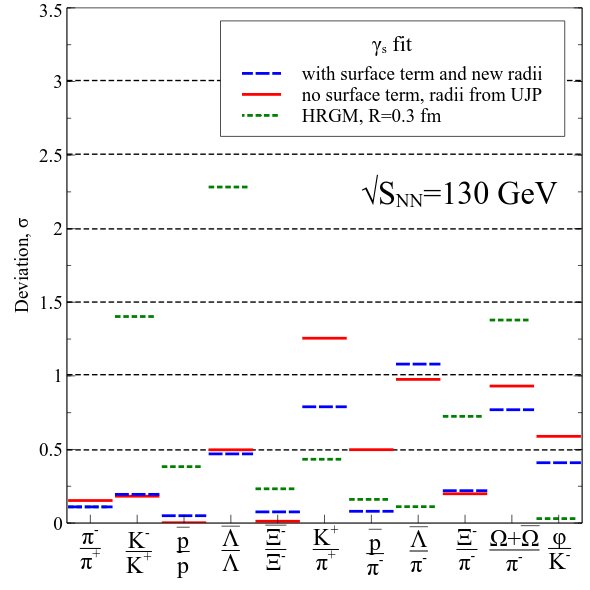}}
\caption{Deviations of theoretically predicted hadronic yield ratios from experimental values in units of
 experimental error $\sigma$ are shown for the center of mass collision energies $\sqrt{s_{NN}} = 8.8$ GeV 
 and $\sqrt{s_{NN}} = 130$ GeV. 
 Dashed lines correspond to the IST EoS fit, while the 
 solid lines correspond to the original HRGM  fit \cite{Sagun}. For a comparison the results obtained by the HRGM1 with a single hard-core  radius $R_{all} = 0.3$ fm for all hadrons are also shown (for more details see text).
}
\label{Bugaev_talk_fig1}       
\end{figure}
\begin{figure}[h]
\centering
\mbox{\includegraphics[width=70mm,clip]{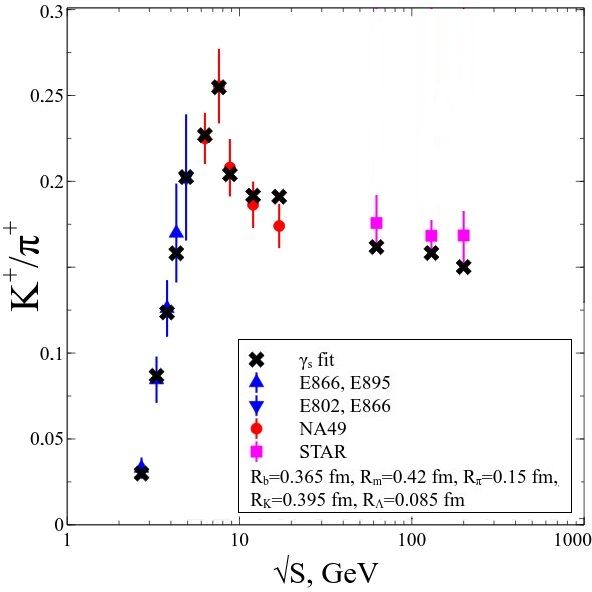}
\hspace*{0.2mm}
\includegraphics[width=70mm,clip]{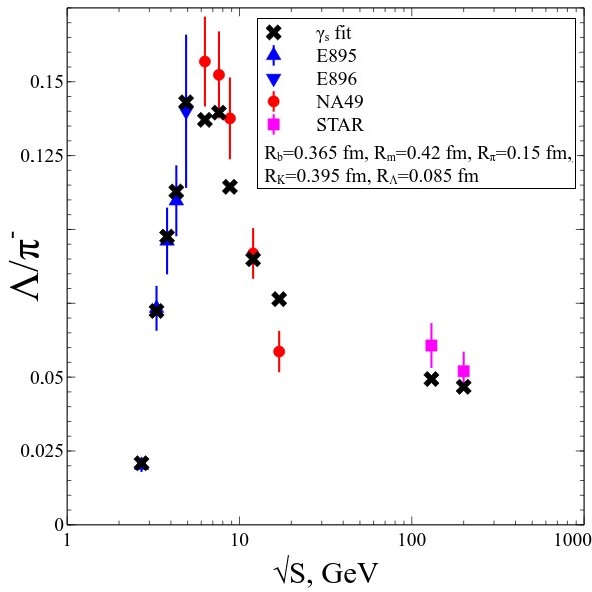}}
\caption{The fit results obtained by the IST EoS: $\sqrt{s_{NN}}$  dependence  of  $K^+/\pi^+$ (left  panel)  and   $\Lambda/\pi^{-}$ (right panel) ratios. For more than a decade these ratios were the most problematic one to reproduce by the HRGM.}
\label{Bugaev_talk_fig2}       
\end{figure}

The parameter  $\alpha =1.25$ was fixed in Refs. \cite{IST1,IST2}, since this value allows us to simultaneously reproduce the third and forth virial coefficients of the gas of classical hard spheres. 
Such a formulation of the IST EoS is  used to simultaneously  fit $111$ independent hadron yield  ratios measured at AGS, SPS and RHIC energies. In this fit the factor $\gamma_s$ and the chemical potentials  $\mu_B$ and $\mu_{I3}$  are regarded  as the free parameters  and  we found that the best description of these data is reached for  the following 
HCR of baryons $R_{b}=0.365 \pm0.03$ fm, mesons $R_{m}=0.42\pm0.04$ fm, pions $R_{\pi}=0.15\pm0.02$ fm, kaons
$R_{K}=0.395 \pm0.03$ fm and $\Lambda$-hyperons $R_{\Lambda}=0.085 \pm0.015$ fm (new radii hereafter).  
These values of the HCR generate $\chi_1^2/dof=57.099/50 \simeq 1.14$ \cite{IST2}.  Some selected  results of this fit are shown 
in Figs. \ref{Bugaev_talk_fig1}, \ref{Bugaev_talk_fig2}.

\begin{figure}[h]
\centering
\mbox{\includegraphics[width=110mm,clip]{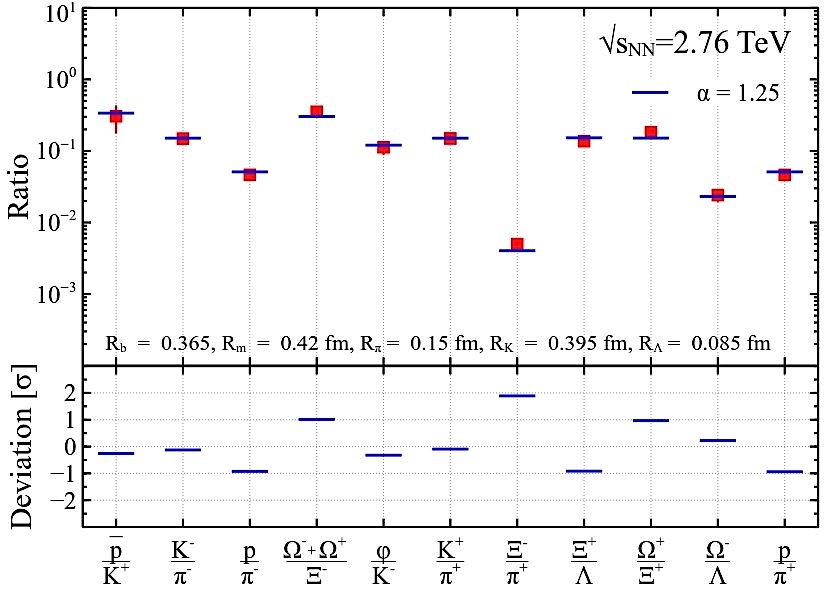}
}
\caption{The results obtained by the IST EOS on fitting  the ALICE data  with the new HCR  found in  \cite{IST1} from fitting the AGS, SPS and RHIC data.  The found   CFO temperature is  $T_{CFO} \simeq 148  \pm 7$ MeV. The  fit  quality is 
 $\chi^2/dof \simeq 8.92/10 \simeq 0.89$. The upper panel shows the fit of the ratios, while the lower panel shows the deviation between data and theory in units of  estimated error.
}
\label{Bugaev_talk_fig4}       
\end{figure}

The found  HCR were fixed and then used  to fit 11 independent hadron yield ratios measured by the ALICE Collaboration (for details see
\cite{IST1,IST2}) 
with a single fitting parameter, namely the CFO temperature, since all the chemical potentials were set to zero, while  the factor $\gamma_s$  was
set to 1.  The fit quality $\chi^2_2/dof \simeq 8.92/10 \simeq 0.89$  of the ALICE data is similar to  the one found for the combined fit of the AGS, SPS and RHIC data (see Fig. \ref{Bugaev_talk_fig4}). Therefore, the combined quality of the AGS, SPS, RHIC and ALICE data description achieved  by the IST EoS  \cite{IST2} is  $\chi^2_{tot}/dof \simeq 66.02/60 \simeq 1.1$. 

In order to show  the importance of the multicomponent hard-core repulsion in Fig.  \ref{Bugaev_talk_fig1} the obtained results are compared to the HRGM with a single HCR of hadrons $R_{all} = 0.3$ fm (HRGM1 hereafter).  The HRGM1 employs  the quantum statistics for all hadrons  and, hence, it is similar to the model of   Ref. \cite{Andronic:05}. The main differences with  Refs. \cite{Andronic:05,Stachel:2013zma} are: 
the HRGM1 includes the widths of all hadronic resonances  for all temperatures and it is used to fit not all hadronic ratios, but only the independent ones. Such a comparison with the multicomponent versions of  HRGM is necessary in order to illustrate the disadvantages  of the one component case compared to the multicomponent formulation. 
The  fit quality obtained by the HRGM1 for AGS, SPS and RHIC energies is $\chi_1^2/dof=75.134/54 \simeq 1.39$ \cite{IST2} which is essentially worse compared to the IST results.  For this case the value of common HCR was not fitted and, hence, the number of degrees of freedom for HRGM1 is 54.
Using the HRGM1 to fit the ALICE data we obtained the fit quality  $\chi^2_2/dof \simeq 12.4/10 \simeq 1.24$ \cite{IST2}.  Hence, the quality of the combined fit for all energies with the HRGM1 is  $\chi^2_{tot}/dof \simeq 87.53/64 \simeq 1.37$, i.e., it is worse than the one found for the multicomponent IST EoS. 

These results  clearly demonstrate  that additional 3 or 4 HCR can, indeed,  essentially improve the quality of the fit of more than hundred independent hadron multiplicity  ratios and, hence, such an improvement provides  a high confidence in the extracted   parameters of CFO. Apparently, this is also a strong argument in favor of $R_N =0.365\pm0.03$ fm found by the IST EoS. Moreover,
from the left panel of Fig. \ref{Bugaev_talk_fig1} one can see that  the proton to negative pion ratio cannot be described within the HRGM1, while it is well described within either of the HRGM  multicomponent  formulations.


\section{Nuclear Matter IST EoS and Proton Flow Constraint}
\label{sect2_Bugaev_talk}
Now we turn to a discussion the quantum version of the IST EoS used to model the nuclear liquid-gas phase transition.  The model pressure $p$ is a solution of the system  ($R_N$ is the HCR on nucleons)
\begin{eqnarray}
\label{I}
p&=&p_{id}(T,\nu_p) - p_{int}\bigl(n_{id}(T,\nu_p)\bigl)\,,
\end{eqnarray}
\begin{eqnarray}
\label{II}
\Sigma&=&R_N\, p _{id}(T,\nu_\Sigma)\,,
\end{eqnarray}
where the grand canonical  pressure $p_{id}(T,\nu)$ and particle number density  $n_{id}(T,\nu)=\frac{\partial p_{id}}{\partial\,  \nu}$  of noninteracting point-like fermions 
are given by the expressions  \cite{ISTU0}
\begin{eqnarray}
\label{III}
p_{id}=Tg_N\int\frac{d^3p}{(2\pi)^3}\ln\left[{\textstyle 1+\exp\left(\frac{\nu-\sqrt{p^2+m^2}}{T}\right)} \right]\,, ~
n_{id}= 
g_N\int\frac{d^3p}{(2\pi)^3} \left[{\textstyle \exp\left(\frac{\sqrt{p^2+m^2}-\nu}{T}\right)+1} \right]^{-1}.
\end{eqnarray}
Here the system temperature  is $T$,  $m_N=940$ MeV is the nucleon mass and  the nucleon degeneracy factor is $g_N=4$.

The term $-p_{int}$ in Eq. (\ref{I}) represents the mean-field contribution to the pressure generated  by an attraction between the nucleons. 
Clearly, the repulsive scattering channels are also present in nuclear matter. However, at densities below $n_{max}\simeq 0.8~fm^{-3}$, which is the maximal density of the flow constraint \cite{Danielewicz}, the repulsion  is  suppressed,  since at  these particle number densities  the mean nucleon separation is larger than  $r_{min}=\left(\frac{3}{4\pi n_{max}}\right)^{1/3}\simeq 0.7$ fm.  But at such distances  the microscopic nucleon-nucleon potential is  attractive \cite{Gross}, whereas  the remaining  repulsive interaction can be safely accounted by the particle hard-core repulsion.

The quantity $\Sigma$ in Eq. (\ref{II}) is a one-component analog of the IST  coefficient of Eq. (\ref{EqII})  first introduced in Ref. \cite{IST14}  in order to distinguish it from the eigensurface tension of ordinary nuclei. Here it is appropriate to explain that  the IST appears  
because the virial expansion of the pressure includes the  terms which are proportional not only to the eigenvolume $V_0=\frac{4\pi}{3}R_N^3$,  but also to the eigensurface $S_0=4\pi R_N^2$ of a particle with the HCR $R_N$  \cite{IST14}. This  surface term contribution  is just accounted by the IST coefficient $\Sigma$. The meaning of $\Sigma$ as the surface tension coefficient can be easily seen from the effective chemical potentials which are related to  the baryonic chemical potential $\mu$ as
\begin{eqnarray}
\label{V}
\nu_p&=&\mu-pV_0-\Sigma S_0+U\bigl(n_{id}(T,\nu_p)\bigl)\,,\\
\label{VI}
\nu_\Sigma&=&\mu-pV_0 -\alpha\Sigma S_0 +U_0\, .
\end{eqnarray}
Here $\Sigma$ is conjugated to $S_0$, while  the attractive mean-field potentials are  denoted as $U\bigl(n_{id}(T,\nu_p)\bigl)$ and $U_0 = const$. From these expressions one can  conclude that the effects  of hard-core repulsion  are only partly accounted by the eigenvolume of particles, while the rest  is determined by  their eigensurface and the IST coefficient $\Sigma$ (for more details see \cite{IST14}). 
Note  that  the presence of the pressure of point-like particles $p_{id}$   in Eqs. (\ref{I}) - (\ref{II}) is a typical feature of  EoSs  formulated in the Grand Canonical Ensemble. 

The system (\ref{I})-(\ref{VI}) is a concrete realization of the quantum model suggested in \cite{IST3}. The self-consistency condition 
\begin{equation}
\label{VII}
p_{int}(n)=n\, U(n)- \int_0^n  dn' \, U(n') \,,
\end{equation}
 relates the interaction pressure $p_{int}\bigl(n_{id}(T,\nu_p)\bigl)$ and the corresponding mean-field potential $U\bigl(n_{id}(T,\nu_p)\bigl)$ and it guarantees the fulfillment of all thermodynamic identities \cite{IST3} for the quantum IST (QIST)  EoS.  

It is necessary to stress   that substituting  the  constant potential $U_0\bigl(n_{id} (T,\nu_\Sigma) \bigr)  = const$ into the consistency condition  (\ref{VII}),  one automatically finds  that  the corresponding  mean-field pressure should vanish, i.e. $\tilde p_{int}\bigl(n_{id}(T,\nu_\Sigma)\bigl)=0$.  
Note  also that different density dependences  of  the attractive mean-field potentials $U\bigl(n_{id})$ and $U_0$ simply 
reflect   the different origins  of their forces. Thus,   $U\bigl(n_{id})$ is generated by the bulk part of interaction, while $U_0$ is related  to the surface part. The meaning of $U_0$ potential can be better understood  after  the non-relativistic  expansion of the nucleon  energy $\sqrt{m^2 + p^2} \simeq m +  \frac{p^2}{2 m}$ staying  in the momentum  distribution function of  Eq. (\ref{III}): $U_0$ lowers  the nucleon mass to the value $m - U_0$ which is similar to  the relativistic mean-field approach.  The particle number density from the usual thermodynamic identity 
\begin{equation}
\label{VIII}
n_N= \frac{\partial p}{\partial \mu} = \frac{n_{id}(T,\nu_p)}{1+V_0~n_{id}(T,\nu_p) + \frac{3\, V_0\,n_{id}(T,\nu_\Sigma) }{1+3(\alpha-1)V_0\, n_{id}(T,\nu_\Sigma)}}\, . 
\end{equation}

To be specific  the power form of the mean-field potential  \cite{ISTU0}
motivated by Ref. \cite{Gorenstein93} 
\begin{eqnarray}
\label{IX}
U(n_N)=C_d^2 n_N^\kappa\, \quad \Rightarrow  \quad  p_{int}(n_N)=\frac{\kappa}{\kappa+1}C_d^2 n_N^{\kappa+1}\, ,
\end{eqnarray}
 is used. 
Here the mean-field contribution to the pressure $p_{int}(n_N)$ is found  from the consistency condition (\ref{VII}). This is one of the simplest choices of the mean-field potential which includes two parameters only, i.e. $C_d^2$ and 
$\kappa$. Since the parameter $\alpha$ is fixed already (see preceding section),  the other two parameters of the 
QIST model are the hard-core radius $R_N$ and the constant potential $U_0$. 
\vspace*{-0.4mm}
\begin{figure}[th]
\centering
\mbox{\includegraphics[width=70mm,clip]{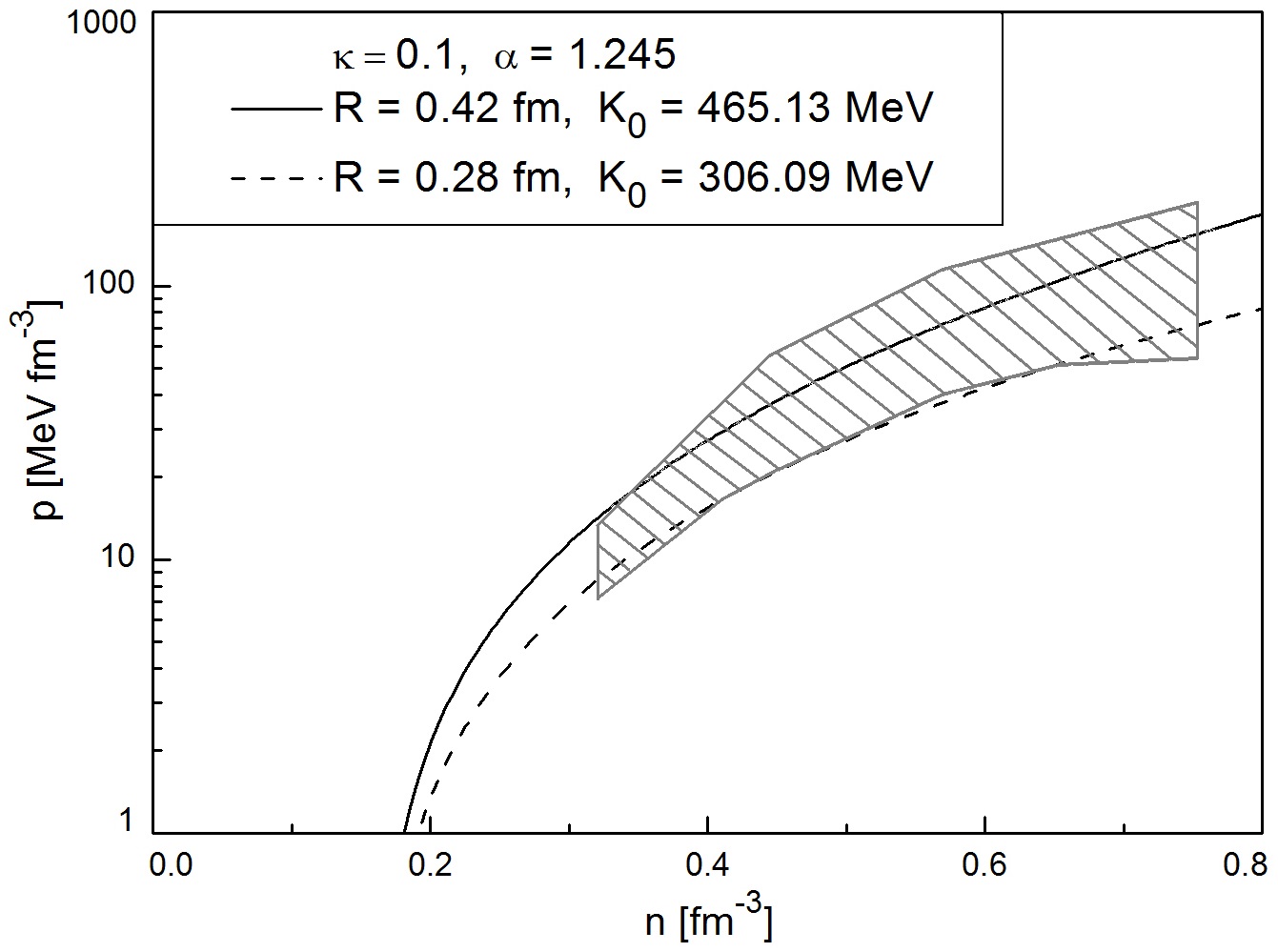}
\hspace*{0.2mm}
\includegraphics[width=73mm,clip]{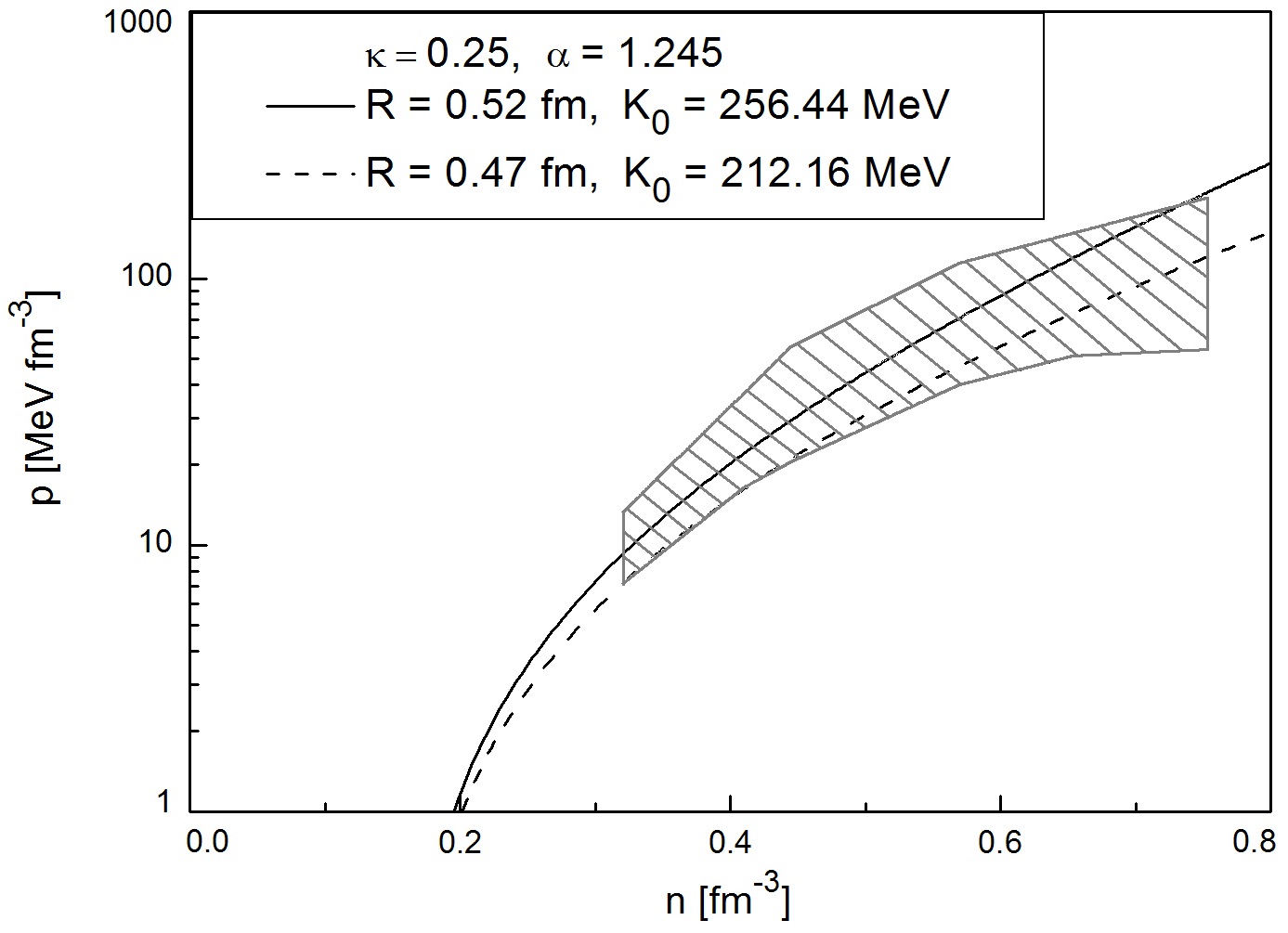}}
\caption{Density dependence of the system pressure is shown for  several set of parameters
which are specified in the legend of each panel. See Table I for more details. 
The dashed area corresponds to the  proton flow constraint of Ref.  \cite{Danielewicz}
}
\label{Bugaev_talk_fig3}
\end{figure}
\begin{table*}[t!]
\begin{tabular}{|c|c|c|c|c|c|c|c|c|c|c|c|c|c|}
                                                                                                      \hline
                                                       & \multicolumn{2}{|c|}{$\kappa=0.1$}
                                                       &\multicolumn{2}{|c|}{$\kappa=0.15$}&\multicolumn{2}{|c|}{$\kappa=0.2$}
                                                       &\multicolumn{2}{|c|}{$\kappa=0.25$}     \\ \hline
                    $R_N~[fm]$                     &  0.28 &  0.42  & 0.35  &  0.48  & 0.41  &  0.50 & 0.47   &  0.52    \\ \hline
 $C_d^2~[MeV\cdot fm^{3\kappa}]$ &284.98&325.06&206.05&229.57&168.15&179.67&146.97&152.00 \\ \hline
                 $U_0~[MeV]$                  &567.32&501.65&343.93&312.83&231.42&217.76&162.03&157.41 \\ \hline
                 $K_0~[MeV]$                  &306.09&465.13&272.55&405.97&242.56&322.80&217.16&256.44 \\ \hline

%
\end{tabular}
\label{table1}
\caption{Different sets of parameters which simultaneously reproduce the properties of normal nuclear matter ($p=0$ and $n=n_0 = 0.16 $fm$^{-3}$ at $\mu=923$ MeV, see text for details) and obey the proton  flow constraint on the nuclear matter EoS along with incompressibility factor $K_0$ and parameters of CEP. $R_N, C_d^2,  U_0$ and $\kappa$ are the adjustable parameters of QIST EoS.}
\end{table*}

The QIST EoS with four adjustable parameters is able  to simultaneously reproduce the main properties of symmetric nuclear matter, i.e.  a vanishing pressure $p_N=0$ at zero temperature $T=0$ and the normal nuclear particle number density $n_0 = 0.16$ fm$^{-3}$ and the value of its binding energy per nucleon $W_0 = \frac{\epsilon_N}{n_N} - m = -16$ MeV  (where $\epsilon_N$ is  the energy density). Hence, the baryonic chemical potential of nucleons is $\mu = 923$ MeV. The QIST  EoS with the attraction term (\ref{IX}) was normalized to these properties of nuclear matter ground state and, simultaneously, it was fitted \cite{ISTU0} to obey the proton flow constraint \cite{Danielewicz}. 
In the present  analysis we consider a few values of parameter $\kappa=0.1,~0.15,~0.2,~0.25$.  For each  value of parameter $\kappa$ the two curves in the $n_N-p$ plane were found in such a way that the upper curve  is located not above the upper branch of the flow constraint, while the lower curve is located not below the lower branch of this constraint. The details are clear from two panels of Fig. \ref{Bugaev_talk_fig3}. 
Notice that this is highly nontrivial result for an EoS with only four adjustable parameters, since to parameterize the proton flow constraint alone one needs at least 8 independent points!  For a comparison we mention that  in  Ref. \cite{Dutra14} it is shown that only 104 of  relativistic mean-field  EoSs out of 263 analyzed in there  are able to obey the proton flow constraint \cite{Danielewicz} despite the fact that they have 10 or even more adjustable parameters. 

However, as was demonstrated in Ref.~\cite{Klahn:2006ir}, the lower bound of the proton flow constraint would correspond to a sequence of neutron stars with a maximum mass of only $\sim 1$ M$_\odot$ and thus would not fulfill the constraint from the observed mass of $2.01\pm 0.04$ M$_\odot$ for pulsar PSR J0348+432 \cite{Antoniadis:2013pzd}.
In Ref.~\cite{Klahn:2006ir} it was also shown that an equation of state which should fulfill the constraint on the maximum mass should follow the upper bound of the flow constraint. 
Thus the IST EoS in the parametrization optimized for explaining particle yields from heavy-ion collisions would be too soft for the phenomenology of neutron stars, i.e. at $T=0$. This was noticed recently in Ref.~\cite{NS17}.

The values of parameter $\kappa$ above 0.33 were not considered, since a good description of the proton flow constraint cannot be achieved for $\kappa \ge 0.33$ \cite{ISTU0}. 
On the other hand, the values of parameter $\kappa$ below 0.1 were not considered  too since  they correspond to unrealistically   large values of the incompressibility constant $K_0 \equiv 9\frac{\partial p}{\partial n_N}\bigl|_{T=0,~n_N=n_0}$.   As one can find  from Table I for  $\kappa=0.1$ the minimal value of the incompressibility constant $K_0$ is about 306 MeV, while for  $\kappa < 0.1$ it is  even larger.  

From Table I one can see that the range of $R_N$ is still wide, i.e. $R_N \in [0.28; 0.52]$ fm.
The QIST EoS, however,  allows one to obtain an essentially narrower  range of  the nucleon HCR $R_N$. Indeed, if one requires that this EoS should be applicable at the maximal value of particle number density $ n_{max} \simeq 0.8$ fm$^{-3}$ of the proton flow  constraint, then such a condition can be written as
\begin{eqnarray}\label{EqX2}
\frac{4}{3} \pi R_N^3 n_{max}~ \le~ \eta_{max} \,.
\end{eqnarray}
Here the range of the QIST EoS applicability is given by the maximal packing fraction $\eta_{max}$ of the model. Assuming that the maximal packing fraction of the QIST EoS  is $\eta_{max} =0.2 $, i.e. it is  similar to the Boltzmann version of the IST EoS \cite{IST1,IST2}, one gets the following inequality on the nucleon hard-core radius $R_N \le 0.4$ fm  and, hence, one finally obtains  $0.28\, {\rm fm} \le R_N \le 0.4$ fm. 

The quantum virial expansion developed in \cite{IST3} both for  the quantum VdW and QIST EoS allows us to obtain even a narrower range of values which is consistent with the S-matrix approach \cite{Pasi18} to the EoS of the gas of nucleons  at temperatures above 100 MeV.  For an extended  discussion see also Ref. \cite{VovchBU}.   In particular, the quantum second virial coefficient $a^S_2(T)$ of a nucleon gas as obtained from realistic S-matrix approach provides  approximately the following inequalities \cite{Pasi18,VovchBU}
\begin{eqnarray}\label{EqXVIII}
0.5 ~{\rm fm}^3 \le  a^S_2(T) \le 1.25~{\rm fm}^3  \quad {\rm for} \quad 100~{\rm MeV} \le T  \le 170~{\rm MeV} \,.
\end{eqnarray}
These inequalities correspond to the conditions $0.31$ fm $\le R_N \le 0.42$ fm, if one uses the classical definition of the HCR. 
It is interesting that these inequalities are similar to the ones found above for the QIST EoS. 
Using the results of  Ref.  \cite{IST3} the  second $a_{2}^{IST}$ and third $a_{3}^{IST}$ virial coefficients for the repulsive part of the QIST EoS for nucleons
can be cast as
\begin{eqnarray}
\label{EqXIX}
a_{2}^{IST} = 4V_0 + a_2^{(0)} \,, \quad
a_{3}^{IST}  \simeq  [16-18(\alpha-1)] V_0^2 +  5 V_0 a_2^{(0)} + a_3^{(0)} \,, \quad 
\end{eqnarray}
where the second $a_2^{(0)}$ and the third $a_3^{(0)}$ virial coefficients of point-like nucleons which in the non-relativistic approximation for fermions can be written as 
\begin{eqnarray}
\label{EqXX}
a_2^{(0)} \simeq  2^{-\frac{5}{2}} \omega_N \simeq 0.177 \omega_N \,, ~
a_3^{(0)}  \simeq 2 \left[ 2^{-4} - 3^{-\frac{5}{2}} \right]  \omega_N^2 \simeq -3.4 \cdot 10^{-3} \omega^2_N  \,, ~ \omega_N = \frac{1}{g_N} \left[ \frac{2 \pi \hbar^2}{T m_N} \right]^\frac{3}{2} \,.
\end{eqnarray}
Introducing an effective second virial coefficient of nucleons $a_2^{eff} (n_N) \equiv a_2^{IST}+ n_N a_{3}^{IST}$ which depends on particle number density of nucleons $n_N$ and assuming that the nucleonic contribution to the  HRGM is given by the repulsive part of the QIST EoS (\ref{I}), one can use the effective second virial coefficient $a_2^{eff} (n_N)$ to constrain the values of  HCR further. 
Our analysis shows that for the nucleon densities below $n_N \simeq 3 n_0 = 0.48$ fm$^{-3}$ the fourth and higher virial coefficients are not important and, hence, we can require that up to this nucleon density the coefficient   $a_2^{eff} (n_N )$ obeys the constraint  (\ref{EqXVIII}). This leads to the follows range 
of $R_N$ values: $R_N \in [0.275; 0.36]$ fm. In other words, for such a range of values of the nucleonic HCR not only the second, but also the third virial coefficient of nucleons will provide the fulfillment of the constraint (\ref{EqXVIII}).


At first glance this result may look surprising, since one does not see any  important role  of the quantum third virial coefficient. 
A close inspection shows that due to the small value of the coefficient which enters the expression for  $a_3^{(0)}$, the quantum effects
are important at temperatures below 20 MeV, while  at $T \ge 100$ MeV  the coefficients  $a_3^{(0)}$ and $a_2^{(0)}$  are small, since 
$\omega_N (T=100 \,{\rm MeV}) \simeq 1$ and it is a decreasing function of 
$T$.  
As a result at  $T \ge 100$ MeV the values of the coefficients $a_{2}^{IST}$ and $a_{3}^{IST}$ are defined by the HCR of nucleons and the parameter $\alpha$.

However, when the QIST EoS is required to simultaneously  fulfill the gravitational mass-radius relation of neutron stars and the proton flow constraint,  one finds somewhat  larger values of the HCR of nucleons, namely $R_N \in [0.42; 0.47]$ fm \cite{NS17}. 
Note that within the recent excluded nucleon volume generalization of the relativistic meanfield model "DD2" by Typel \cite{Typel:2016srf}
even larger values of the HCR of nucleons were used in the description of neutron star phenomenology such as mass-radius relations \cite{Kaltenborn:2017hus}, moment of inertia \cite{Alvarez-Castillo:2018pve}, tidal deformabilities \cite{Paschalidis:2017qmb,Alvarez-Castillo:2018pve} and cooling \cite{Grigorian:2016leu}.
The "DD2\underline{ }p40" EoS used in these works would correspond to a nucleon HCR of $R_N=0.62$ fm which is at the very limit of what is compatible with the recent constraint on the compactness of neutron stars stemming from the gravitational wave signal measured for the inspiral phase of the neutron star merger GW170817
\cite{Annala:2017llu}.    
These results indicate that the repulsive core of the nucleon-nucleon interaction
depends on the properties of the medium since
the description of static neutron star properties at zero temperature require a stiffer EoS than the one which is successfully reproducing the hadronic multiplicities measured in HIC. What is the physical  reason for such a difference? 

It has been demonstrated that the repulsive part of effective density-dependent interactions of the Skyrme type (e.g., the one by Vautherin and Brink \cite{Vautherin:1971aw}) can be reproduced by the quark exchange interaction between nucleons (quark Pauli blocking) \cite{Ropke:1986qs} in analogy to the hard-sphere model of molecular interactions which is based on the electron exchange interaction among atoms (see, e.g., Ebeling et al. \cite{Ebeling:2008mg}) which is captured, e.g., in the Carnahan-Starling EoS \cite{Carnahan}.
The repulsive Pauli blocking effect between composite particles is especially pronounced at low temperatures, in the regime of quantum degeneracy.
 

Note that the QIST  model  offers a simple way to make a stiffer EoS at higher pressures or densities. Actually, as it was mentioned in the first paper on IST EoS \cite{IST14},  the parameter $\alpha$ may, in principle, be a function which depends on the system pressure. 
Therefore, it would be interesting to generalize  the QIST EoS  and to include into it the pressure or density dependence of  the parameter $\alpha$ 
and/or of  the HCR of nucleons. Then having more adjustable parameters and adding  more astrophysical constraints as, e.g., for an upper limit on the maximum mass as well as lower and upper bounds on the neutron star radius from the binary neutron star merger, one could aim at a best possible description including the proton flow constraint and to find a realistic functional dependence of $\alpha$ and $R_N$ on density and temperature. 
In this respect we would like to mention the possibility to model the excluded nucleon volume in a density and temperature dependent way, even changing the sign so that also attractive interactions are accessible. In this form, Typels excluded volume model \cite{Typel:2016srf} has been used to obtain an equation of state and phase diagram with a second critical endpoint (CEP) beyond the gas - liquid one \cite{Typel:2017vif}. This could be used to mimic effects of the nuclear-to-quark matter phase transition in the QCD phase diagram. 
Within the IST approach the IST coefficient $\Sigma$ stands for attraction effects and therefore the interplay of attraction and repulsion as captured in the (medium dependent) parameters $\alpha$ and $R_N$, respectively, could eventually lead to similar behaviour and a second CEP in the phase diagram.  

\section{Conclusions}

Here we thoroughly discussed the IST approach to model the EoS of hadronic and nuclear matter and analyzed different constraints on the HCR of nucleons. The most successful formulation of HRGM gives us $R_N\simeq 0.365\pm0.03$ fm, while the QIST EoS of nuclear matter leads to $R_N\simeq 0.34\pm0.06$ fm. At the same time a comparison of quantum virial coefficients with with S-matrix approach gives us $R_N\simeq 0.32\pm0.04$ fm. Therefore, the most probable range of HCR of nucleons which is consistent with different constraints following from the hadronic and nuclear matter properties  is $R_N \in [0.3; 0.36]$ fm. Since  applications of the QIST EoS to the neutron star properties require somewhat larger HCR of nucleons 
\cite{NS17}, we conclude that the QIST EoS for neutron stars should be improved further, especially an interaction between nucleons 
at  high  particle number densities which are typical for the neutron stars core.   The generalized  QIST EoS which considers the density and temperature dependence of the parameters $\alpha$  and $R_N$
may provide a very effective way to solve this problem. 

\acknowledgments{ 
The authors are thankful to 
I. N. Mishustin, R. D. Pisarski and S. A. Moszkowski for fruitful discussions and valuable comments. 
The work of K.A.B., 
A.I.I., V.V.S., B.E.G., D.O.S. and G.M.Z. was supported by 
the grants  launched by the Section of Nuclear Physics of National Academy of Sciences  of Ukraine. 
V.V.S. thanks the 
Funda\c c\~ao para a Ci\^encia e Tecnologia (FCT), Portugal, for the
financial support through the Grant No.
UID/FIS/00099/2013 to make research at the Centro de Astrof\'{\i}sica e 
Gravita\c c\~ao (CENTRA), Instituto Superior T\'ecnico, Universidade de 
Lisboa. 
The work of L.V.B. and E.E.Z. was supported by the Norwegian 
Research Council (NFR) under grant No. 255253/F50 - CERN Heavy Ion 
Theory. 
L.V.B. and K.A.B. thank the Norwegian Agency for International Cooperation and Quality Enhancement in Higher Education for financial support, grant 150400-212051-120000 «CPEA-LT-2016/10094 From Strong Interacting Matter to Dark Matter».
D.B.B. is grateful to the COST Action CA15213 ``THOR" for networking support and 
to  the MEPhI Academic Excellence program grant No 02.a03.21.0005 for partial support. 
D.B.B. and L.T. acknowledge support from the Polish National Science Centre (NCN) under grant no. UMO-2014/13/B/ST9/02621.
The work of A.V.T. was partially supported by the Ministry of Science and Education of the Russian Federation, grant No. 3.3380.2017/4.6, and by National Research Nuclear University ``MEPhI'' in the framework of the Russian Academic Excellence Project (contract no. 02.a03.21.0005, 27.08.2013).  The work of A.I.I. was done within
the project SA083P17 of Universidad de Salamanca launched by the Regional Government of Castilla y Leon and the European Regional Development Fund. 
}

%

\begin{thebibliography}{90}
%
%

\bibitem{Horst86}
H. St\"ocker  and W. Greiner, Phys. Rep. {\bf 1986}, 137, 227

\bibitem{Lattimer12}
 J. M.  Lattimer,  
 Annu. Rev. Nucl. Part. Sci.  {\bf  2012}, 62, 485  and references therein.

\bibitem{StellarSMM13}
 N. Buyukcizmeci, A. S. Botvina, I. N. Mishustin,  
  Astrophys.\ J.\  {\bf 789}, 33 (2014).
  
 \bibitem{David15}
%
{S. Benic, D. Blaschke, D. E. Alvarez-Castillo, T. Fischer and S. Typel},
{Astron. Astrophys.} {\bf 2015}, 577, A40 
  
\bibitem{Dutra14} 
M. Dutra et al., 
 Phys. Rev. C {\bf 2014}, 90,  055203 and references therein.

\bibitem{Stone:2014wza} 
J.~R.~Stone, N.~J.~Stone and S.~A.~Moszkowski,
Phys.\ Rev.\ C {\bf 2014}, 89, 044316





\bibitem{Andronic:05}
A. Andronic, P. Braun-Munzinger and J. Stachel, Nucl. Phys. A {\bf 2006}, 772, 167


 \bibitem{Oliinychenko:12}
D. R. Oliinychenko, K. A. Bugaev and  A. S. Sorin, Ukr. J. Phys.  {\bf 2013}, 58,  211    
  
\bibitem{HRGM:13}
K. A. Bugaev, D. R. Oliinychenko, A. S. Sorin, G. M. Zinovjev, Eur. Phys. J. A {\bf 2013}, 49, 30

\bibitem{SFO:13}
K. A. Bugaev {\it et al.,}
Europhys. Lett. {\bf 2013}, 104, 22002 
 
 \bibitem{Sagun}
V. V. Sagun, Ukr. J Phys. {\bf 2014}, 59,  755

\bibitem{Sagun2}
%
V. V. Sagun {\it et al.,}
Ukr. J. Phys. {\bf 2014}, 59, 1043

\bibitem{Stachel:2013zma}
  J.~Stachel, A.~Andronic, P.~Braun-Munzinger and K.~Redlich,
  J.\ Phys.\ Conf.\ Ser.\  {\bf 2014}, 509, 012019 and references therein.
 
\bibitem{BugaevUJP16}
%
K. A. Bugaev {\it et al.,} Ukr. J. Phys.  {\bf 2016}, 61, 659 and references therein


\bibitem{Bugaev:2015}
  K.~A.~Bugaev {\it et al.,}
 Phys. Part. Nucl. Lett. {\bf 2015}, 12, 238

\bibitem{Bugaev:2016a}
  K.~A.~Bugaev  {\it et al.,}
  Eur. Phys. J. A {\bf 2016}, 52, 175; and Eur. Phys. J. A {\bf 2016}, 52, 227

   

\bibitem{Bugaev:2017a} 
  K.~A.~Bugaev
 {\it et al.,}
 Phys. Part. Nucl. Lett. {\bf 2018}, 15, 210
 
\bibitem{IST1}
%
K. A. Bugaev {\it et al.,}
Nucl. Phys. A {\bf 2018}, 970, 133

\bibitem{IST2}
%
V. V. Sagun  {\it et al.,}
Eur. Phys. J. A {\bf 2018}, 54, 100

\bibitem{IST3}
%
K. A. Bugaev, A. I. Ivanytskyi, V. V. Sagun, E. G. Nikonov and G. M. Zinovjev,
arXiv:1704.06846 [nucl-th] (to appear in Ukr. J Phys. (2018)) and references therein

  
\bibitem{Raju}
%
{R. Venugopalan and M. Prakash}, 
{Nucl. Phys. A} {\bf 1992}, 546, {718}     

\bibitem{Satarov10}
%
L. M. Satarov,  M. N. Dmitriev and I. N. Mishustin,
Phys. Atom. Nucl.  {\bf 2009}, 72,  1390 

\bibitem{KABkinFO1}
%
K. A. Bugaev,
Nucl. Phys. A {\bf 1996}, 606, 559

\bibitem{KABkinFO2}
%
K. A. Bugaev,
Phys. Rev. Lett. {\bf 2003}, 90, 252301 and references therein

\bibitem{Hagedorn}
%
R. Hagedorn, Nuovo Cim. Suppl.  {\bf 1965}, 3, 147

\bibitem{RHIC17}
%
L. Adamczyk {\it et al.,} [STAR Collaboration],
Phys. Rev. C {\bf 2016}, 93, 021903

\bibitem{NICA}
%
P.  Senger,
{Eur. Phys. J. A} {\bf 2016}, 52, 217 and references therein

\bibitem{FAIR}
%
P.  Senger,
{Nucl. Phys. A} {\bf 2011}, 862-863, 139 and references therein

\bibitem{Rafelski}
J. Rafelski, Phys. Lett. B {\bf 1991}, 62, 333

\bibitem{IST14}
%
V. V. Sagun, A. I. Ivanytskyi, K. A. Bugaev and I. N. Mishustin,
Nucl. Phys. A {\bf 2014}, 924, 24


\bibitem{ISTU0}
A. I. Ivanytskyi, K. A. Bugaev, V. V. Sagun, L. V. Bravina and E. E. Zabrodin, 
Phys. Rev. C {\bf 2018}, 97,  064905
 
\bibitem{Danielewicz}
P. Danielewicz, R. Lacey and W. G. Lynch,  Science {\bf 2002}, 298, 1593 

\bibitem{Gross}
F. Gross, J. W. Van Orden and K. Holinde, Phys. Rev. C {\bf 1992}, 45, 2094 and references therein


\bibitem{Gorenstein93}
M. I. Gorenstein {\it et al.,} J. Phys. G {\bf 1993}, 19, 69

\bibitem{Klahn:2006ir} 
T.~Klahn {\it et al.},
Phys.\ Rev.\ C {\bf 2006}, 74, 035802

\bibitem{Antoniadis:2013pzd}
J.~Antoniadis {\it et al.},
Science {\bf 2013}, 340, 1233232	 

\bibitem{NS17}
%
V. V. Sagun and  I. Lopes,
Astrophys. J {\bf  2017}, 850, 75


\bibitem{Pasi18}
P. Huovinen and P. Petreczky,
Phys. Lett. B {\bf 2018}, 777, 125

\bibitem{VovchBU}
V. Vovchenko, A. Motornenko, M. I. Gorenstein and H. Stoecker,
Phys.\ Rev.\ C {\bf 2018}, 97, 035202

\bibitem{Typel:2016srf} 
S.~Typel,
Eur.\ Phys.\ J.\ A {\bf 2016}, 52, 16

\bibitem{Kaltenborn:2017hus} 
M.~A.~R.~Kaltenborn, N.~U.~F.~Bastian and D.~B.~Blaschke,
Phys.\ Rev.\ D {\bf 2017}, 96, 056024

\bibitem{Alvarez-Castillo:2018pve} 
D.~E.~Alvarez-Castillo, D.~B.~Blaschke, A.~G.~Grunfeld and V.~P.~Pagura,
arXiv:1805.04105 [hep-ph].

\bibitem{Paschalidis:2017qmb} 
V.~Paschalidis, K.~Yagi, D.~Alvarez-Castillo, D.~B.~Blaschke and A.~Sedrakian,
Phys.\ Rev.\ D {\bf 2018}, 97, 084038


\bibitem{Grigorian:2016leu} 
H.~Grigorian, D.~N.~Voskresensky and D.~Blaschke,
Eur.\ Phys.\ J.\ A {\bf 2016}, 52, 67

\bibitem{Annala:2017llu} 
E.~Annala, T.~Gorda, A.~Kurkela and A.~Vuorinen,
Phys.\ Rev.\ Lett.\  {\bf 2018}, 120, 172703

\bibitem{Vautherin:1971aw} 
D.~Vautherin and D.~M.~Brink,
Phys.\ Rev.\ C {\bf 1972}, 5, 626

\bibitem{Ropke:1986qs} 
G.~R\"opke, D.~Blaschke and H.~Schulz,
Phys.\ Rev.\ D {\bf 1986}, 34, 3499

\bibitem{Ebeling:2008mg} 
W.~Ebeling, D.~Blaschke, R.~Redmer, H.~Reinholz and G.~R\"opke,
J.\ Phys.\ A {\bf 2009}, 42, 214033

\bibitem{Carnahan}
N. F. Carnahan and K. E. Starling, J. Chem. Phys. {\bf 1969}, 51, 635

\bibitem{Typel:2017vif} 
S.~Typel and D.~Blaschke,
Universe {\bf 2018}, 4, 32

\end{thebibliography}
%
\reftitle{References}

\end{document}